\documentstyle[12pt]{article}
\@addtoreset{equation}{section} \makeatother

\newcommand{\be}{\begin{equation}}
\newcommand{\ee}{\end{equation}}
\newcommand{\bee}{\begin{eqnarray}}
\newcommand{\beee}{\begin{array}}
\newcommand{\eee}{\end{eqnarray}}
\newcommand{\eeee}{\end{array}}

\newcommand{\f}{\frac}
\newcommand{\ptl}{\partial}
\newcommand{\p}{\partial}
\newcommand{\nn}{\nonumber}
\newcommand{\half}{\frac{1}{2}}

\newcommand{\ga}{\alpha}
\newcommand{\gb}{\beta}

\newcommand{\gvep}{\varepsilon}

\newcommand{\bw}{{\bar{w}}}

\newcommand{\dc}{{\dot{c}}}

\newcommand{\da}{{\dot{a}}}
\newcommand{\db}{{\dot{b}}}

\newcommand{\dga}{{\dot{c}}}
\newcommand{\de}{{\dot{e}}}

\newcommand{\Hh}{{ N}}

\newcommand{\M}{{\cal M}}

\renewcommand{\S}{{\cal S}}

\begin{document}

\begin{flushright}
 FIAN/TD/01--06\\
\end{flushright}\vspace{1.5cm}

\begin{center}
{\large\bf
  Higher Spin Conformal Currents in Minkowski Space
} \vglue 0.6  true cm \vskip1cm

O.A. Gelfond$^1$, E.D. Skvortsov$^2$  and M.A.~Vasiliev$^2$ \vglue 0.3  true cm

${}^1$Institute of System Research of Russian Academy of Sciences,\\
Nakhimovsky prospect 36-1, 117218, Moscow, Russia
\vglue 0.3  true cm

${}^2$I.E.Tamm Department of Theoretical Physics, Lebedev Physical Institute,\\
Leninsky prospect 53, 119991, Moscow, Russia
\vskip1.5cm
\end{center}

\begin{abstract}
\baselineskip=24pt Using unfolded formulation of free equations for
massless fields of all spins we obtain explicit form of gauge
invariant higher-spin conformal conserved charges bilinear in $4d$
massless fields of arbitrary spins.

\end{abstract}

\section{Introduction}\label{Generalities}

In this note we give an explicit form of gauge invariant higher-spin
(HS) conserved currents built of $4d$ massless fields of all spins.
 To the best of our knowledge a
realization of the conformal HS currents built of massless fields
of all spins in the $4d$  Minkowski space has not been yet
available in the literature in the full generality, although some
particular examples of conformal HS currents built of massless
fields of lower spins $s\leq 1$ were considered. In particular,
$x$-independent HS conformal currents built of massless scalar,
spinor and Maxwell field were found in \cite{Ans} and
$x$-dependent HS currents built of massless scalar and spinor were
found in \cite{KVZ}. We extend these results in two directions: we
allow (i) the constituent fields  to carry any spin and (ii)
explicit dependence on the space-time coordinates.

Our construction is based on the unfolded formulation of dynamical
equations in the form of zero-curvature equations \cite{Ann} and is
analogous to the construction of HS currents \cite{cur} in the
generalized space-time with matrix coordinates
\cite{mat,Sorokin,BHS, Mar}.

\section{Unfolded\,\, Massless\,\, Field\,\, Equations\,\, in\,\, $4d$\,\,
Minkowski\,\, space}

In \cite{Ann,BHS} it was shown that the equations for (field
strengths of) massless fields of all spins in Minkowski space can
be concisely formulated in the unfolded form \bee \label{minun}
\f{\p}{\p x^{a\db}}C(w,\bw|x) +\f{\p^2}{\p w^a\p
\bw{}^{\db}}C(w,\bw|x)=0\,. \eee Here $w^a$ and $\bw^\db$ are
auxiliary commuting conjugated two-component spinor coordinates
($a,b = 1,2$ and $\dot{a},\dot{b} = \dot{1}, \dot{2}$) and $x
^{a\db}$ are Minkowski coordinates in two-component spinor
notations. The two-component indices are raised and lowered as
follows \be A^a = \gvep^{ab}A_b\,,\qquad A_a =
\gvep_{ba}A^b\,,\qquad \gvep_{ab} = -\gvep_{ba}
\,,\quad\gvep_{12}=1 \ee and analogously for dotted indices. The
relationship with the tensor notations is based on \be V^{a\db}=
A^\nu \sigma_\nu^{a\db}\,,\qquad \ee
where $\sigma_\nu^{a\db}$
($\nu = 0,1,2,3$)  are four Hermitian $2\times 2$ matrices.

The meaning of the equation (\ref{minun})  is as follows.
The fields $C(w,\bw|x)$ are assumed to be expandable in power series
in $w^a$ and $\bw^\da$
\be
C(w,\bw|x)=\sum\limits_{m,n=0}^{\infty} C_{a_1\dots a_n\,\da_1\dots \da_m}(x)w^{a_1}\dots w^{a_n}
\bw^{\da_1}\dots \bw^{\da_m}\,.
\ee
The operator
\bee\nn
\Hh_{w,\bw}=w^a\f {\p}{\p w^a }-\bw{}^\da\f{ \p}{\p \bw{}^\da} \eee
commutes with  $\f{ \p^2}{\p w^a \p \bw{}^\da}$. Solutions of the
equation (\ref{minun}) with fixed eigenvalues of $\Hh_{w,\bw}$ form
invariant subspaces which describe fields of different helicities
$h$ \be
 \Hh_{w,\bw}
  C(w,\bw|x)= 2h C(w,\bw|x)\,.
\ee
The meaning of the fields $C(w,\bw|x)$  is as follows  \cite{BHS}.
 The holomorphic fields
\bee\nn
C(w,0|x) =
\sum\limits_{2s=0}^{\infty}
C_{a_1\dots a_{2s}}(x) w^{a_1}\cdots w^{a_{2s}}
\eee
 and their complex conjugates\footnote{$w^a$ is complex conjugated to $\bw^{\da}$}
\bee\nn
C(0,\bar{w}|x) =
\sum\limits_{2s=0}^{\infty}
 C_{\da_1\dots \da_{2s}}(x) \bw^{\da_1}\cdots \bw^{\da_{2s}}
\eee
 describe, respectively, selfdual (positive helicity) and
antiselfdual (negative helicity) gauge invariant on-mass-shell
nontrivial combinations
of derivatives of massless gauge fields of all spins
$s=0,1/2,1,\ldots \infty$,  where
\bee  \nn
 w^a\f{\p}{\p w^a} C(w,0|x)= 2s C(w,0|x)\,
\eee
and
\bee \nn
 \bw^{\dot{a}}\f{\p}{\p \bw^{\dot{a}}} C(0,\bw|x)= 2s C(0,\bw|x)\,.
\eee
 These include scalar ($s=0$)
\bee  \nn
 c(x) =C(0,0|x)\,,
\eee spinor ($s=1/2$) \bee   \nn c_a (x)= \f{\p}{\p w^a}
C(w,0|x)\Big |_{w=0}\,,\qquad \bar{c}_\da (x)= \f{\p}{\p
\bw^{\dot{a}}} C(0,\bw|x)\Big |_{\bw=0}\,, \eee
 Maxwell tensor ($s=1$)
\bee  \nn
c_{ab} (x)= \frac{1}{2} \frac{\p^2 }{\p w^a w^b} C(w,0|x)\Big |_{w=0}\,,\qquad
\bar{c}_{\da\db} (x)=  \frac{1}{2}  \f{\p^2}{\p \bw^\da \p \bw^{\db}} C(0,\bw|x)\Big |_{\bw=0}\,,
\eee
Rarita-Schwinger field strength ($s=3/2$)
\bee \nn
c_{abc} (x)= \f{1}{3!}
\f{\p^3 }{\p w^a \p w^b \p w^c} C(w,0|x)\Big |_{w=0}\,,\quad\,\,
\bar{c}_{\da\db\dot{c}} (x)= \f{1}{3!}\f{\p^3}{\p \bw^{\dot{a}}
\p \bw^{\dot{b}}\p\bw^{\dot{c}}} C(0,\bw|x)\Big |_{\bw=0}\,,
\eee
Weyl tensor ($s=2$)
\bee\nn
c_{abcd} (x)&=& \f{1}{4!}
\f{\p^4 }{\p w^a \p w^b \p w^c \p w^d} C(w,0|x)\Big |_{w=0}\,,\\ \nn
\bar{c}_{\da\db\dot{c}\dot{d}} (x)&=& \f{1}{4!}\f{\p^3}{\p \bw^{\dot{a}}
\p \bw^{\dot{b}}\p\bw^{\dot{c}} \p\bw^{\dot{d}}} C(0,\bw|x)\Big |_{\bw=0}\,,
\eee
etc.

The {\it primary} fields are those contained in $ C(w,0|x)$ and
their complex conjugates $C(0,\bw|x)$. These are lowest order gauge
invariant combinations of derivatives of massless gauge fields,
which turn out to be of order $[s]$ for a spin $s$ field and were
considered by many authors (see e.g. \cite{wein}). The descendants
are described by those components of $C(w,\bw |x)$ that depend both
on $w$ and on $\bw$ and therefore are expressed in terms of
derivatives of the primary fields by (\ref{minun}).

The dynamical HS field equations are the following consequences of
(\ref{minun}) \be
\label{Dynamical}
\f{\p}{\p x^{a\db}}\f{\p}{\p w_a}C(w,0|x)= 0,    \qquad
\f{\p}{\p x^{a\db}}\f{\p}{\p \bw_\db}C(0,\bw|x)= 0
\ee for $s\neq 0 $ and the massless Klein-Gordon equation \bee
\label{mineq} \f{\p^2}{\p x^{a\db}  \p x_{a\db}  }c(x)=0 \eee
for  $s=0$ scalar (for $s>0$ it is a consequence of
(\ref{Dynamical})).

Given function $C(w,\bw|0)$ of the spinors $w^a$ and $\bw^\da$ it
uniquely reconstructs a solution of the equation (\ref{minun}) by
\bee \nn C(w,\bw|x) = \exp{\left (-x^{a\dot{b}}\f{\p^2}{\p w^a \p
\bw^\db }\right )} C(w,\bw|0)\,. \eee Other way around, given
solution of the equations (\ref{minun}) the full dependence on $w$
and $\bar{w}$ is reconstructed as follows. The Taylor expansion
gives \bee \nn C(w,\bar{w}|x) = \exp\left(w^a\f{\p}{\p v^a}
+\bw^{\da} \f{\p}{\p \bar{v}^\da}\right) C(v,\bar{v}|x)\Big
|_{v=\bar{v}=0}\,. \eee For a given helicity $h\geq 0$ we obtain
\bee \nn C(w,\bar{w}|x) = \f{1}{(2h)!}\left(w^b\f{\p}{\p
v^b}\right)^{2h} F_h \left(w^a\f{\p}{\p v^a} \bw^{\da} \f{\p}{\p
\bar{v}^\da}\right) C(v,\bar{v}|x)\Big |_{v=\bar{v}=0}\,, \eee where
the function \bee \nn F_h (r) =
\sum_{n=0}^{\infty}\f{(2h)!}{n!(2h+n)!}(r)^n \eee is related to the
regular Bessel functions $I_{k}(x)$
 (see, e.g., \cite{bessel}) as follows
\bee \nn
 \f{r^{h}}{(2h)!}F_h (r)= I_{2h}(2r^{\half}).
\eee Now, using again  equation (\ref{minun}), we obtain for a
field with a positive helicity $h$ \bee \label{SelfDual}
C^h(w,\bar{w}|x) =F_h (-w^a \bw^{\db}\p_{a \db} )C^h(w,0|x)\,,
\eee where $\p_{a\da} = \f{\p}{\p x^{a\da}}$. Analogously, one
obtains for negative helicities $h<0$ \be \label{AntiSelfDual}\nn
C^h(w,\bar{w}|x) =F_{|h|} (-w^a \bw^{\db} \p_{a \db})
C^h(0,\bw|x)\,. \ee This reconstructs the dependence on $w$ and
$\bw$.

\section{Higher-Spin Conformal Currents}
 From the equation (\ref{minun}) it follows \cite{BHS} that the
 field equations for massless fields of all spins are $sp(8)$
symmetric with $sp(8)$ being a maximal finite-dimensional
subalgebra of the infinite-dimensional  HS symmetry.
This symmetry is conformal because $sp(8)$
contains the $4d$ conformal algebra $su(2,2)$ as a subalgebra.
 The infinite set of
conformal HS symmetries suggests the existence of conserved HS
currents.

The HS charges in Minkowski space should have the form \bee
\label{Q3} Q(\eta) = \int\limits_{\Sigma^{3}} \Omega^{3}(\eta)
\,, \eee
where
$\eta$ denotes the HS symmetry parameters,
$\Omega^{3}(\eta) $ is a on-mass-shell closed $3$-form dual
to the conserved current, and $\Sigma^{3}$ is an arbitrary
$3$-dimensional surface in the Minkowski space-time usually
identified with the space surface $R^3$, {\it i.e.} the Cauchy
surface for the problem.

Using the unfolded form of the massless field equations it is easy
to write down explicit formulae for  the
conserved HS charges in $4d$ Minkowski space.  Let us consider the following
3-form  in  Minkowski space $M^4$
\bee \label{clo3}
\Omega^3(\eta)=
\eee
$$
 dx_{a\da} \wedge dx^{a\dc} \wedge
dx^{c\da} \, \eta_{ b_{1} \ldots b_{l}\, \db_{l+1} \ldots \db_{t}\,
}{}^{\alpha_1 \ldots \alpha_s}
x^{b_{1} \de_1}\ldots x^{b_{l}\de_l} x^{e_{l+1}\db_{l+1}} \dots x^{e_t\db_{t}}
T_{c\dc \, \alpha_1\ldots \alpha_s\, e_{l+1} \ldots e_{t}\, \de_{1}
\ldots \de_{l}\, }\,,
$$
where $\eta_{\beta_1\ldots\beta_t}{}^{\alpha_1 \ldots \alpha_s}$
are arbitrary HS symmetry parameters symmetric in lower and upper
indices, and the generalized stress tensor
$T_{\alpha_1\ldots\alpha_n}$ is also symmetric. We use notation
with four-component Greek indices being equivalent to a pair of
dotted and undotted two-component indices, e.g., $\alpha = a,\da$.

The form (\ref{clo3}) is closed
 \be
\label{cl3}
 d\,
\Omega^3(\eta)=0\, \ee provided that the generalized stress tensor
$T_{\alpha_1 \ldots \alpha_{n}}(x)$ satisfies the conservation
condition \bee \label{qeq} \frac{\ptl}{\ptl x^{b \db}}\, {T}^{b
\db}{}_{ \alpha_1\ldots \alpha_{n-2}}(x)=0\,. \eee Indeed, taking
into account that the generalized stress tensor $T_{\alpha_1 \ldots
\alpha_{n}}(x)$ is symmetric in its indices, (\ref{cl3})  is a
simple consequence of the fact that \bee\nn dx_{a\da} \wedge
dx^{a}{}_{\dga} \wedge dx_{c}{}^{\da}\wedge dx^{b\dot{b}} \f{\p}{\p
x^{b\dot{b}}} =\f{1}{4} dx_{a\da} \wedge dx^{a}{}_{\db} \wedge
dx_{b}{}^{\da}\wedge dx^{b\dot{b}}
 \f{\p}{\p
x^{c\dot{c}}}\,.
\eee

For the case with an equal number of dotted and undotted indices
among the indices $\alpha$ in  (\ref{qeq}),
it amounts to the usual  conservation condition for traceless symmetric
tensors which is well-known to be related to conformal HS symmetries \cite{Ans}.
The equation (\ref{qeq}) tells us however that, in the general case, all
irreducible tensors of the $4d$ Lorentz algebra may appear as generalized
HS conserved stress tensors except for those described by purely dotted
(i.e., selfdual) or
purely undotted (i.e., antiselfdual) components.
Note that, in the tensorial language,
generalized stress tensors of integer spins  are described by various
traceless Lorentz tensors that have symmetry properties of Young tableaux
with at most two rows. The components of $T_{\alpha_1 \ldots \alpha_{n}}(x)$
that do not contribute to the conserved charge are described by various
two-row rectangular Young tableaux.

The key observation is that the generalized stress tensor \bee
\label{TCCM} T^{k\,l}_{\alpha_1 \ldots \alpha_{n}}(x)= \f{\p}{\p
y^{\alpha_1}}\dots \f{\p}{\p y^{\alpha_n}} \Big (C^k (y |x) C^l (i
y|x)\Big ) \Big|_{y=0}\,, \eee where  $y^\alpha =(w^a , \bw^\da
)$, satisfies the conservation condition (\ref{qeq}) provided that
the  field $C^k (y|x^{a\da})$ satisfies the $4d$ unfolded equation
(\ref{minun}). Indeed, from (\ref{minun}) it follows \bee \nn
\f{\p}{\p x^{b\db}} \f{\p}{\p w_b}\f{\p}{\p \bw_\db} \Big (C^k (y
|x) C^l (i y|x)\Big ) =\\ -\nn \f{\p}{\p w_b}\f{\p}{\p \bw_\db}
\left(C^k (y |x) \f{\p}{\p w^b}\f{\p}{\p \bw^\db} C^l (i y|x)
-(\f{\p}{\p w^b}\f{\p}{\p \bw^\db} C^k (y |x))  C^l (i y|x)\right
) =0\,. \eee

Note that
the conserved currents built of HS fields according to (\ref{TCCM})
contain higher derivatives. This is in agreement with the analysis of
\cite{des} as well as with the general
property of HS theories that their interactions contain higher
derivatives \cite{pos,FV}.

\section{Examples}

In this section we consider some examples of conserved currents
resulting from the general construction.

In terms of two-component fields, the dynamical equations
(\ref{Dynamical}), (\ref{mineq}) on the (anti)selfdual components
$c_{a_1 a_2 \ldots a_{2s}}(x)$ and $\bar{c}_{\dot{a}_1 \dot{a}_2
\ldots \dot{a}_{2s}}(x)$ read
 \be
\label{dirhs}
 \p^{a_1\dot{a_1}}c_{a_1 a_2 \ldots
a_{2s}}=0, \qquad \p^{a_1\dot{a_1}}\bar{c}_{\dot{a}_1 \dot{a}_2
\ldots \dot{a}_{2s}}=0.
\ee
These equations  imply that space-time
derivatives of the field strengths are symmetric in dotted and
undotted indices separately.

The straightforward substitution of (\ref{SelfDual}),
(\ref{AntiSelfDual}) into ({\ref{TCCM}}) gives  the generalized
stress tensor that contains $p$ derivatives acting on spin-$s$
selfdual and spin-$s^\prime$ antiselfdual constituent fields \be
\label{GC}
T_{a(2s+p),\dot{a}(2s'+p)}^{s,s',p|k,l}=\sum_{j=0}^{j=p}(-)^j
\frac{(2s)!(2s+p)!(2s')!(2s'+p)!}{(2s+j)!(p-j)!(2s'+p-j)!j!}
\p_{a\dot{a}(j)}c_{a(2s)}^k\p_{a\dot{a}(p-j)}\bar{c}^l_{\dot{a}(2s')}\,,
\ee where we use  notation
 \be
\p_{a\dot{a}(k)}\equiv\frac{\p^k}{\p x^{a\dot{a}}\ldots \p
x^{a\dot{a}}}\,, \ee and the indices denoted by the same letter are
assumed to be symmetrized (with the convention that the
symmetrization is a projector, {\it i.e.} the repeated
symmetrization leaves a symmetrized tensor unchanged). Analogously
one can construct  selfdual-selfdual
$T^{s,s',p|k,l}_{a(2s+2s'+p),\dot{a}(p)}$ and
antiselfdual-antiselfdual $T^{s,s',p|k,l}_{a(p),\dot{a}(2s+2s'+p)}$
generalized stress tensors.

The generalized irreducible angular momentum tensors obtained from
(\ref{clo3}) have the form \bee
&M^{s,s',p|m,m',n|k,l}_{a(2s+p-m+m'-n),\dot{a}(2s'+p+m-m'-n)}=\\&=T_{a(2s+p-m-n)b(m)c(n),\dot{a}(2s'+p-m'-n)\dot{b}(m')\dot{c}(n)}^{s,s',p|k,l}
{{x^{b(m)}}_{\dot{a}(m)}}{{x^{\dot{b}(m')}}_{a(m')}}{x^{c\dot{c}(n)}}\,,\eee
with ${x^{b(m)}}_{\dot{a}(m)}\equiv \overbrace{{x^b}_{\dot{a}}\ ...\
{x^b}_{\dot{a}}}^{m}$.

For the particular case of fields of equal spins, we obtain the
generalized stress tensors \be \label{genstr}
T_{a(2s+p),\dot{a}(2s+p)}^{s,s,p|k,l}=\sum_{j=0}^{j=p}
     \frac{(-)^j ((2s)!)^2((2s+p)!)^2}{(2s+j)!(p-j)!(2s+p-j)!j!}
\p_{a\dot{a}(j)}c_{a(2s)}^k(x)\p_{a\dot{a}(p-j)}\bar{c}^l_{\dot{a}(2s)}
\ee and $T^{s,s,p|k,l}_{a(4s+p),\dot{a}(p)}$,
$T^{s,s,p|k,l}_{a(p),\dot{a}(4s+p)}$ corresponding to rank $2s+p$
symmetric traceless tensors $T_{\mu(2s+p)}^{s,s,p|k,l}$ in the
tensor notation.

Let us consider some lower-spin examples. A spin-0 massless scalar
field $c^k(x)$ satisfies the Klein-Gordon equation \be
\p^{\mu}\p_\mu c^k(x)=0, \ee equivalent to  (\ref{mineq}). The HS
totally symmetric conserved currents built of higher derivatives of
the scalar field \cite{Ans,KVZ} are
 \be T_{a(p),\dot{a}(p)}^{0,0,p|k,l}=\sum_{j=0}^{j=p}
(-)^{j}\left\{
\frac{p!}{(p-j)!j!}\right\}^2\p_{a\dot{a}(j)}c^k(x)\p_{a\dot{a}(p-j)}
c^l(x)\,. \ee For the particular cases of $p=1$ and $p=2$ we obtain
the  electric
 current
 \be
J^{k,l}_{a,\dot{a}}=\p_{a\dot{a}}c^l c^k-\p_{a\dot{a}}c^k c^l\ee and
the improved stress tensor \be T_{aa,\dot{a}\dot{a}}^{0,0,2|k,l}=
c^k\p_{a\dot{a}}\p_{a\dot{a}}c^l-4\p_{a\dot{a}}c^k\p_{a\dot{a}}
c^l+\p_{a\dot{a}}\p_{a\dot{a}}c^k c^l\,, \ee which is symmetric in
the (discarded) color indices. In tensor notation, these have the
form
 \be \label{chiral} J^{k,l}_\mu=c^k\p_\mu
c^l-c^l\p_\mu c^k \ee and \be \label{ScalarEM} T_{\mu\nu}=\p_\mu
c\p_\nu c-\frac12 c \p_{\mu\nu} c-\frac14 \eta_{\mu\nu}\p_\lambda
c \p^\lambda c\,. \ee

A spin-$\frac12$ field $\psi(x)$ satisfying the massless Dirac equation
$ \gamma_\mu\p^{\mu}\psi(x)=0 $
is described by $c_a (x)$ and
$\bar{c}_{\dot{a}}(x)$ that satisfy (\ref{dirhs}).
Neglecting the color indices, the electric current
 $T^{\frac12,{\frac12},0}$ and stress tensor  $T^{\frac12,{\frac12},1}$ are
 \be
T^{\frac12,{\frac12},0}_{a,\dot{a}}=c_a \bar{c}_{\dot{a}}\,, \ee
\be T^{\frac12,{\frac12},1}_{a(2),\dot{a}(2)}= 2\Big ( c_a
\p_{a\da} \bar{c}_{\dot{a}}- \p_{a\da} c_a \bar{c}_{\dot{a}}  \Big
)\,. \ee

A supercurrent, which mixes spin-0 and spin-$\frac12$ fields,
is given by $T^{\frac12,0,1}$,
 \be
T^{\frac12,0,1}_{aa,\dot{a}}= 2c_a\p_{a\dot{a}} c-\p_{a\dot{a}} c_a
c\, \ee and its complex conjugate.

A massless spin-1 field, can be described  by a
gauge invariant field strength satisfying the
Maxwell equations
\be \p^\mu
F_{\mu\nu}=0\,,\qquad \p_{[\rho}
F_{\mu\nu]}=0\,.
\ee
In terms of two-component spinors,
$F_{\mu\nu}$ is described by $c_{aa}$ and $\bar{c}_{\dot{a}\dot{a}}$,
while  the Maxwell equations have the form (\ref{dirhs}).

The stress tensor \be T_{\mu\nu}=-{F_\mu}^\sigma
F_{\sigma\nu}+\frac14\eta_{\mu\nu}F^2\ee is described by
$T^{1,{1},0}$ \be T^{1,{1},0}_{aa,\dot{a}\dot{a}}=4
c_{aa}\bar{c}_{\dot{a}\dot{a}} \,.
\ee
Analogously to the scalar field
case, there exist totally symmetric HS conserved currents built of
higher derivatives of the spin-1 field strength \cite{Ans}. These
are the generalized stress tensors  $T^{1,{1},p}$.

With the aid of the stress tensor $T_{\mu\nu}$ one can construct
an angular momentum tensor \be
M_{\mu,\nu\lambda}=T_{\mu\nu}x_\lambda-T_{\mu\lambda}x_\nu\ee in
the case of spin-0 the corresponding spinor-tensors are
$M^{0,0,2|1,{0},0}_{a,\dot{a}(3)}$,
$M^{0,0,2|0,{1},0}_{a(3),\dot{a}}$ and
$M^{0,0,2|0,{0},1}_{a,\dot{a}}$.

A massless spin-2 field describes the linearized gravity.
The linearized gauge invariant combinations of derivatives of a
linearized metric tensor are given by the linearized Riemann tensor.
Its trace part is zero by virtue of Einstein equations.
The nonzero traceless part is called Weyl tensor $H_{\mu\nu\lambda\rho}$.
As a consequence of Einstein equations, the Weyl tensor satisfies
differential restrictions by virtue of
Bianchi identities. In terms of two-component spinors  Weyl tensor is
 described by the selfdual component $c_{abcd}(x)$ and antiselfdual component
$\bar{c}_{\dot{a}\dot{b}\dot{c}\dot{d}}(x)$.
The consequences of Einstein equations have  the  form (\ref{dirhs}).

It is well-known that there is a conserved current called
Bel-Robinson tensor \cite{Bel}, \cite{Deser} that is  bilinear
in  Weyl tensor. In terms of tensors, it has the form
\be
T_{\mu\nu\lambda\rho}=H_{\mu\sigma\nu\eta}{{{H_\lambda}^\sigma}_\rho}^\eta+
{}^*H_{\mu\sigma\nu\eta}{}^*{{{H_\lambda}^\sigma}_\rho}^\eta,\ee
where
 the Hodge star $*$ denotes the dualization by virtue of the
Levi-Civita tensor $\varepsilon_{\mu\nu\lambda\rho}$. In terms of
two component spinors the Bel-Robinson tensor  is described by
$T^{2,{2},0}$, having the simple form
 \be
T^{2,{2},0}_{a(4),\dot{a}(4)}=
(4!)^2c_{a(4)}\bar{c}_{\dot{a}(4)}\,. \ee Higher symmetric
generalized strength tensors built of the Weyl tensor are given by
the formula (\ref{genstr}) with $s=2$.

\section{Conclusion}

Although the obtained list of conserved currents is infinite it does
not reproduce some of the expected symmetry generators and as such
is  incomplete. This is not surprising because even ordinary
conserved currents like stress tensor and electric charge for higher
spins are not in the presented class of gauge invariant currents.
Indeed, it is well-known that the energy-momentum conservation in
gravity is described in terms of gauge non-invariant pseudo-tensor
\cite{Landau} which, however
  gives rise to gauge invariant total
energy and momentum conservation laws at the level of free fields.
The same happens for all higher spins \cite{des,DeserA,DeserB}. The
reason is simply that the gauge invariant tensors $C(w,\bar{w}|x)$
contain at least $s$ derivatives of a spin-$s$ gauge potential and
therefore can only appear in the conserved currents which themselves
carry sufficiently high spins.

The system of higher spin fields of all spins is  $sp(8)$ invariant.
\cite{mat,BHS}. The pattern of the symmetry parameters and
corresponding conserved currents in two-component spinor notations
is as follows: generalized translations have symmetry parameters
  $\eta^{a\dot{a}}$, $\eta^{aa}$, $\eta^{\dot{a}\dot{a}}$ and
conserved currents
    $T^{k,l}_{a,\dot{a}\dot{a}\dot{a}}$,
    $T^{k,l}_{aaa,\dot{a}}$,
    $T^{k,l}_{aa,\dot{a}\dot{a}}$.
Generalized  Lorentz boosts and dilatation have symmetry parameters
  ${\eta_{b}}^{a}$, ${\eta_{\dot{b}}}^{a}$,
${\eta_{b}}^{\dot{a}}$, ${\eta_{\dot{b}}}^{\dot{a}}$ and conserved currents
   $T^{k,l}_{aa,\dot{a}\dot{b}}x^{b\dot{b}}$, $T^{k,l}_{aab,\dot{a}}x^{b\dot{b}}$, $T^{k,l}_{a,\dot{a}\dot{a}\dot{b}}x^{b\dot{b}}$,
  $T^{k,l}_{ab,\dot{a}\dot{a}}x^{b\dot{b}}$.
  Generalized special conformal transformations have symmetry parameters
 $\eta_{bb}$, $\eta_{b\dot{b}}$, $\eta_{\dot{b}\dot{b}}$ and
conserved currents
 $T^{k,l}_{a,\dot{a}\dot{b}\dot{b}}x^{b\dot{b}}x^{b\dot{b}}$,
  $T^{k,l}_{ab,\dot{a}\dot{b}}x^{b\dot{b}}x^{b\dot{b}}$, $T^{k,l}_{abb,\dot{a}}x^{b\dot{b}}x^{b\dot{b}}$.

The list of generators of this type (which includes the generators
of the usual conformal algebra $su(2,2)\subset sp(8)$)  that can be
constructed in terms of invariant higher spin tensors is quite short
 \be\label{11} T^{k,l}_{aa,\dot{a}\dot{a}}=
c^k\p_{a\dot{a}}\p_{a\dot{a}}c^l-4\p_{a\dot{a}}c^k\p_{a\dot{a}}
    c^l+\p_{a\dot{a}}\p_{a\dot{a}}c^k c^l\;;
    4c_{aa}\bar{c}_{\dot{a}\dot{a}}\;; 2\Big ( c_a\p_{a\da} \bar{c}_{\dot{a}}- \p_{a\da} c_a \bar{c}_{\dot{a}}
    \Big)\,,\ee
    \be \label{12}
T^{k,l}_{aaa,\dot{a}}=2\Big ( c_a\p_{a\da} \bar{c}_{\dot{a}}- \p_{a\dot{a}} c_a \bar{c}_{\dot{a}}
    \Big)\;;\quad6c^k_{aaa}\bar{c}^l_{\dot{a}}\,,
    \ee
    \be\label{13}
    T^{k,l}_{a,\dot{a}\dot{a}\dot{a}}=6\bar{c}^l_{\dot{a}\dot{a}\dot{a}}c^k_{a}\;;\quad
    6 \bar{c}^l_{\dot{a}\dot{a}}\p_{a\dot{a}} c^k -2\p_{a \dot{a}}\bar{c}^l_{\dot{a}\dot{a}} c^k\ee
and obviously incomplete because the $sp(8)$ symmetry mixes fields of
all spins while the generators (\ref{11}), (\ref{12}) and (\ref{13})
do not contain higher spin fields at all. It remains to be
investigated whether it is possible to complete the list of higher spin
conserved currents presented in this paper by the higher spin
pseudotensors which may not be gauge invariant but
allow for the construction of invariant conserved charges.

Finally let us note that the formula for conformal HS currents
presented in this paper is analogous to the formula of
\cite{cur,tens2} for conserved HS  currents in the
ten-dimensional space-time $\M_4$ suggested for the description of
$4d$ massless HS fields in \cite{mat,BHS,Mar}. In fact, the expression
(\ref{TCCM}) for $T^{k\,l}_{\alpha_1 \ldots \alpha_{n}}(y|x^{a\da})$
is the reduction to the Minkowski space
 of the generalized stress tensor \cite{cur}
$T^{k\,l}_{\alpha_1 \ldots \alpha_{n}}(y|X^{\ga\gb})$, where
 $X^{\ga\gb}$ are symmetric matrix coordinates of $\M_4$.
 The conservation condition (\ref{qeq}) is also
the reduction of the conservation condition in $\M_4$. The explicit
relationship between the two constructions, which  requires an
appropriate integration over a noncontractible cycle  in $M_4$
remains to be elaborated, however.

\section*{Acknowledgments}
The work was supported in part by grants RFBR No. 05-02-17654, LSS No.
1578.2003-2 and INTAS No. 03-51-6346.

\end{document}